\documentclass[prb,twocolumn]{revtex4}

\usepackage{graphicx}
\usepackage{amssymb}

\begin{document}

\title{Simulation of the physical properties of the chalcogenide glass
As$_2$S$_3$ using a density-functional-based tight-binding method}

\author{S. I. Simdyankin}
\email{sis24@cam.ac.uk}
\affiliation{Department of Chemistry, University of Cambridge, 
Lensfield Road, Cambridge CB2 1EW, United Kingdom} 

\author{S. R. Elliott}
\affiliation{Department of Chemistry, University of Cambridge, 
Lensfield Road, Cambridge CB2 1EW, United Kingdom}

\author{Z. Hajnal, T. A. Niehaus, Th. Frauenheim}

\affiliation{Fachbereich 6~---~Theoretische Physik, Universit\"at Paderborn, 
             Warburger Stra{\ss}e 100, D-33098, Paderborn, Germany}

\date{\today}

\begin{abstract}
We have used a density-functional-based tight-binding method in order
to create structural models of the canonical chalcogenide glass,
amorphous (a-)As$_2$S$_3$.
The models range from one containing defects that are both chemical
(homopolar bonds) and topological (valence-alternation pairs) in
nature to one that is defect-free (stoichiometric).
The structural, vibrational and electronic properties of the simulated
models are in good agreement with experimental data where available.
The electronic densities of states obtained for all models show clean
optical band gaps.
A certain degree of electron-state localization at the band edges is
observed for all models, which suggests that photoinduced phenomena in
chalcogenide glasses may not necessarily be attributed to the
excitation of defects of only one particular kind.
\end{abstract}

\maketitle

\section{Introduction}

Amorphous chalcogenides (particularly the sulfides, selenides and
tellurides) exhibit intriguing physical properties that are not
observed in their crystalline counterparts.
Some of these unusual properties are extensively used in electronic
and photonic devices \cite{Nalwa_HAEPMD} and there are many potential
applications.

Perhaps the most interesting opto-electronic behavior of these
materials is the metastable structural changes resulting from the
absorption of near-bandgap light \cite{Kolobov_PIMAS}.
The microscopic changes in the atomic structure involved are not
generally observable directly, but they are reflected in measurable
optical, electronic, and mechanical properties \cite{Lee_90,Lowe_86}.
A fully consistent microscopic theory accounting for the
opto-electronic behavior in chalcogenide glasses, however, is still
lacking.
It is therefore of great interest to employ computer simulations in
order to generate and study structural models of such materials.
Provided that these computer-generated models compare well with
available experimental data, they could allow one to monitor the
photo-induced structural changes in the greatest possible detail at
the microscopic level.
Needless to say, such detailed information is not generally presently
available from experiments.

In order to study opto-electronic effects, one needs to perform
quantum-mechanical calculations that are very time consuming.
This imposes a severe constraint on the accessible system size of the
simulated models.
It is possible, however, that small samples may be sufficient to
capture much of the interesting photoinduced behavior due to a high
degree of localization of the photo-excited electron-hole pairs
\cite{Fritzsche_2000,Ovshinsky_2000}.
A few \textit{ab initio} studies of amorphous chalcogenides, paying
particular attention to electronic properties, have been performed by
Drabold and co-workers (see, e.g.,
Refs.~\onlinecite{Cobb_96,Drabold_PRB_00,Li_PRL_2000}).
Although their results agree well with experimental data, the defect
concentration in these structural models is much greater than is
estimated in experiments.
Some of the possible reasons for this are the level of approximation
in the \textit{ab initio} approach employed in these studies and/or
the model preparation history, which may have resulted in too many
quenched-in defects due to very rapid cooling of the sample from the
liquid state.

In this paper, we use a density-functional-based tight-binding (DFTB)
method \cite{Frauenheim_2002} in order to generate and analyze several
models of amorphous diarsenic trisulphide (a-As$_2$S$_3$) with a
controlled and systematic change in the defect concentration.
Apart from being widely regarded as the canonical chalcogenide glass,
this particular material was chosen for analysis for the following
reasons.
First, we have obtained reliable high-quality neutron-scattering
structural data for this material \cite{Lee_98,Haar_2000}.
Second, the properties of a-As$_2$S$_3$ are expected to be similar to
those of a-As$_2$Se$_3$, amorphous diarsenic triselenide, that have
been studied theoretically in Ref.~\onlinecite{Drabold_PRB_00}, and it
is of interest to corroborate this similarity in independent
simulations.
Third, the necessary input DFTB data for sulphur have been previously
generated and extensively tested  \cite{Niehaus_01}.

Although the DFTB method is semiempirical, it allows one to improve
upon the standard tight-binding description of interatomic
interactions by including a DFT-based self-consistent second order in
charge fluctuation (SCC) correction to the total energy.
The flexibility in choosing the desired accuracy while computing the
interatomic forces brings about the possibility to perform much faster
calculations when high precision is not required, and refine the
result if needed.

\section{Methodology}

\subsection{DFTB}
\label{sec:dftb}

The SCC-DFTB model is derived from density-functional theory (DFT) by
a second-order expansion of the DFT total energy functional with
respect to the charge-density fluctuations $\delta n'=\delta
n(\vec{r}\;\!')$ around a given reference density $n'_0 =
n_0(\vec{r}\;\!')$:

\begin{eqnarray}
\label{finerg}
E & = &  \sum_i^{occ} \langle \Psi_i | \hat{H}^0 | \Psi_i \rangle \nonumber \\ 
  & + & \frac{1}{2}\int\!\!\!\int' \left(
\frac{1}{| \vec{r}-\vec{r}\;\!' |}
+\left. \frac{\delta^2 E_{xc}}{\delta n \, \delta n'}
\right|_{n_0} \right)
\delta n \, \delta n' \\
& - &\frac{1}{2}\int\!\!\!\int' \! \frac{n_0' n_0}{|\vec{r}-\vec{r}\;\!'|}
  + E_{xc}[n_0] - \int V_{xc}[n_0]n_0 + E_{ii}, \nonumber 
\end{eqnarray}
where $\int d\vec{r}\;\!'$ is expressed by $\int'$.  Here, $\hat{H}^0
= \hat{H}[n_0]$ is the effective Kohn-Sham Hamiltonian evaluated at
the reference density and the $\Psi_i$ are Kohn-Sham orbitals.
$E_{xc}$ and $V_{xc}$ are the exchange-correlation energy and
potential, respectively and $E_{ii}$ is the core-core repulsion
energy.

To derive the total energy of the SCC-DFTB method,
the energy contributions in Eq. (\ref{finerg}) are further
subjected to the following approximations: \\
1) The Hamiltonian matrix elements $\langle \Psi_i | \hat{H}^0 |
\Psi_i \rangle$ are represented in a minimal basis of confined,
pseudoatomic orbitals $\phi_{\mu}$,
$$\Psi_i = \sum_{\mu} c_{\mu}^i \phi_{\mu}. $$
To determine the basis
functions $\phi_{\mu}$, we solve the atomic DFT problem by adding an
additional harmonic potential $(\frac{r}{r_0})^2 $ to confine the
basis functions \cite{Porezag_95}.  The Hamiltonian matrix elements in this
LCAO basis, $H_{\mu \nu}^0$, are then calculated as follows. The
diagonal elements $H_{\mu \mu}^0$ are taken to be the atomic
eigenvalues and the non-diagonal elements $H_{\mu \nu}^0$ are
calculated in a two-center approximation:
$$
H_{\mu \nu}^0 = <\phi_{\mu} | \hat{T} + v_{eff}[n_{\alpha}^0 + n_{\beta}^0] | \phi_{\nu} >
\;\; \mu \epsilon \alpha, \nu \epsilon  \beta,
$$
which are tabulated, together with the overlap matrix elements $S_{\mu \nu}$
with respect to the interatomic distance $R_{\alpha \beta}$.
$v_{eff}$ is the  is the effective Kohn-Sham potential and
$n_{\alpha}^0$ are the densities of the neutral atoms $\alpha$. \\
2) The charge-density fluctuations $\delta n$ are written as
a superposition of atomic contributions $\delta n_{\alpha}$,
$$
\delta n = \sum_{\alpha} \delta n_{\alpha},$$
which are
approximated by the charge fluctuations at the atoms $\alpha$, $\Delta
q_{\alpha} = q_{\alpha} - q_{\alpha}^0$.  $q_{\alpha}^0$ is the number
of electrons of the neutral atom $\alpha$ and the $q_{\alpha}$ are
determined from a Mulliken-charge analysis.  The second derivative of
the total energy in Eq. (\ref{finerg}) is approximated by a function
$\gamma_{\alpha \beta}$, whose functional form for $\alpha \ne \beta$
is determined analytically from the Coulomb-interaction of two
spherical charge distributions, located at $R_{\alpha }$ and
$R_{\beta}$. For $\alpha = \beta$ it represents the electron-electron
self-interaction on atom $\alpha$. \\
3) The remaining terms in Eq. (\ref{finerg}),
$E_{ii}$ and the energy contributions, which
depend on $n_0$ only, are collected in 
a single energy contribution $E_{rep}$.
$E_{rep}$ is then approximated as a sum of
short-range repulsive potentials,
$$ E_{rep} = \sum_{\alpha \ne \beta} U[R_{\alpha \beta}],$$
which depend on the interatomic distances $R_{\alpha \beta}$.

With these definitions and approximations, the SCC-DFTB
total energy finally reads:
\begin{equation}
\label{etot}
E_{tot} = \sum_{i \mu \nu} c_{\mu}^i  c_{\nu}^i H_{\mu \nu}^0 +
\frac{1}{2} \sum_{\alpha \beta} \gamma_{\alpha \beta} \Delta q_{\alpha}
\Delta q_{\beta} + E_{rep}.
\end{equation}
Applying the variational principle to the energy functional (\ref{etot}),
one obtains the corresponding Kohn-Sham equations:
\begin{eqnarray}
  \label{kse}
    \sum_\nu c_{\nu i} (H_{\mu\nu} - \epsilon_i S_{\mu\nu}) =
  0 , \;\;\; \forall\;\mu ,\:i\;\\
\nonumber H_{\mu\nu} =\langle\phi_\mu| {H_0}|\phi_\nu\rangle+
  \frac{1}{2}
  S_{\mu\nu}\sum_{\zeta}(\gamma_{\alpha\zeta}+\gamma_{\beta\zeta})\Delta
  q_\zeta, 
\end{eqnarray}
which have to be solved iteratively for the wavefunction expansion
coefficients $c_{\mu}^i$, since the Hamiltonian matrix elements depend on
the $c_{\mu}^i$ due to the Mulliken charges.  Analytic first
derivatives for the calculation of interatomic forces are readily
obtained, and second derivatives of the energy with respect to atomic
positions are calculated numerically.

The repulsive pair potentials $U[R_{\alpha \beta}]$ are constructed by
subtracting the DFT total energy from the SCC-DFTB electronic energy
(first two terms on the right-hand side of eq.(\ref{etot})) with
respect to the bond distance $R_{\alpha \beta}$ for a small set of
suitable reference systems.

To sum up, in order to parameterize the method for a new element, the
following steps have to be taken. First, DFT calculations have to be
performed for the neutral atom to determine the LCAO basis functions
$\phi_\mu$ and the reference densities $n^0_\alpha$. Here the
confinement radius can in principle be chosen different for the
density ($r_0^n$) and each type of atomic orbital ($r_0^{s,p,d}$). We
usually take $r_0$ to be the same for s- and p-functions. In a minimal
basis, this yields a total number of two adjustable parameters for
elements in the first and second rows, while there are three if
d-functions are included. After this, the different matrix elements can
be calculated and the pair potentials $U[R_{\alpha \beta}]$ are
obtained as stated above for every combination of the new element with
the ones already parameterized.

In this study, we used the same tabulated data set for sulphur-sulphur
interactions as in Ref.~\onlinecite{Niehaus_01}, and the As-As, As-S
and S-As data sets were generated according to the procedure outlined
above.
The confinement radii for the As pseudoatomic densities (r$_d$ =
9.8~a.u.  and r$_w$ = 4.5~a.u.), as well as the As-As repulsive
pair-potential were determined in accord with other ongoing efforts
related to GaAs systems.
The cage-like As$_4$S$_6$ molecule was used to calculate the As-S
repulsive pair potential, so that after finding the minimum energy
configuration of the molecule in all-electron DFT-LDA calculations
using the NRLMOL program \cite{NRLMOL}, a regular scaling of the As-S
bond-lengths was performed, keeping the overall T$_d$ symmetry.
Then the acquired potentials were tested on other clusters, such as
As$_2$S and As$_4$S$_4$ molecules, with an overall good agreement of
the binding energies and configurations between SCC-DFTB and the
reference all-electron DFT-LDA NRLMOL results.
When these data sets were used in order to optimize the geometry of
the crystal structure of orpiment (c-As$_2$S$_3$) in SCC-DFTB, the
agreement with the experimental structure \cite{Mullen_72} was within
2\%.

\subsection{Preparation of structural models}
\label{sec:prep}

In experiments, bulk glasses are usually prepared from the melt by
rapidly cooling (quenching) the sample.
Although it appears impossible to achieve experimentally realistic
cooling rates in molecular-dynamics computer simulations, some
empirical procedures result in models that can be in good agreement
with experiments.
In order to prepare realistic models of a-As$_2$S$_3$, we use an
algorithm akin to that used, e.g., in
Refs.~\onlinecite{Cobb_96,Drabold_PRB_00,Blaineau_2003}.

The structural model of a-As$_2$S$_3$ was obtained in the course of an
$NVT$ (constant number of particles, $N$, volume, $V$, and
temperature, $T$) molecular-dynamics simulation with periodic boundary
conditions.
Since we are not interested in statistical properties of thermal
fluctuations, the temperature was controlled simply by scaling the
velocities of the constituent particles every few time steps with the
time intervals between the scalings taken randomly with a mean value
of 10 time steps.
We used a time step of 100 a.u. $\approx$ 2.4 fs (1 a.u. =
2.4189$\times10^{-17}$ s) and the Verlet algorithm in order to
integrate the equations of motion.

The starting configuration was a system of 200 (80 arsenic and 120
sulphur) atoms in a cubic supercell, with a side length of
17.25~{\AA}, obtained by rescaling a crystalline configuration of
orpiment (monoclinic, space group 14, P12$_1$/n1) with
$1\times2\times5$ 20-atom unit cells.
The crystalline coordinates were obtained from the Inorganic Crystal
Structure Database and correspond to those reported in
Ref.~\onlinecite{Mullen_72}.
In order to obtain a cubic supercell, we approximated the monoclinic
unit cell of orpiment by an orthorhombic one simply by neglecting the
small deviation of the angle $\beta=90.68^{\circ}$ from the right
angle.
Then we used the experimental glass density of $\rho=3.186$ g/cm$^3$
from Ref.~\onlinecite{Iwadate_99} in order to obtain the side length
of the cubic supercell $L=(N/\rho)^{1/3}=$17.25~{\AA} and the
coordinates of atoms from the cuboid with dimensions 11.48 by 19.15 by
21.28~{\AA}~ were scaled by 1.5, 0.9, and 0.81 in the $x$, $y$, and $z$
directions, respectively.
The use of the crystalline initial configuration gives the correct
stoichiometry and the rescaling of a non-cubic supercell in order to
obtain a cubic one serves the goal of eliminating possible
anisotropies in physical properties.
Note that in Ref.~\onlinecite{Drabold_PRB_00}, the authors cut a cubic
supercell from a crystalline phase of As$_2$Se$_3$ isostructural to
orpiment, thus achieving the above two goals, but disrupting the
periodicity with respect to the periodic boundary conditions.
After thorough equilibration in the simulated liquid state, however,
the use of either of these two prescriptions should lead to models
with similar statistical properties.

The initial configuration was melted and equilibrated first at
$T=3000$~K for 3 ps, and then the resulting configuration was allowed
to equilibrate at $T=1000$~K for 12 ps.
The equilibration criterion used was the convergence of the average
potential energy to a constant value.
At such high temperatures, the accuracy of the calculations appears to
be least significant for the subsequent generation of low-temperature
structural models.
Therefore, at this stage, a minimal basis set of only s and p orbitals
on both the As and S atoms was used, and the tight-binding scheme
of Sec.~\ref{sec:dftb} was used without the self-consistent charge
(SCC) correction in order to speed up the calculations.
It was necessary, however, to use the SCC correction during the
initial 1.5 ps of the $T=3000$~K run, since the large forces resulting
from the presence of small interatomic distances in the distorted
starting configuration otherwise led to numerical instability.

While preparing computer-generated models of glasses, it is customary
to mimic real experiments by reducing the temperature over time
intervals whose length, however, is limited by the available computer
time.
We found that, due to the unrealistically small length of such time
intervals, this approach is rather impractical.
Instead, we used the available computer time to perform an annealing
run at one fixed temperature which is low enough for the process of
the bond-network formation to be activated and high enough for the
topologically connected network to grow sufficiently rapidly.
First, we performed a run corresponding to 6 ps at $T=700$~K with the
SCC correction and the minimal (sp) basis.
Keeping in mind that the simulation was done at constant volume, this
temperature was chosen to be somewhat above the melting temperature of
orpiment at atmospheric pressure ($T_m=592$~K according to
Ref.~\onlinecite{Feltz_AIMG}).
We empirically found that annealing the configuration for the
following 6 ps at a higher temperature of $T=800$~K slightly improved
the quality of the network by increasing the fraction of heteropolar
bonds in the model.
During these two runs, we used a smaller time step of 50 a.u. (1.2 fs). 
Finally, the temperature was nearly instantaneously reduced to
$T=300$~K, quenching the system within a metastable basin on the
potential-energy hypersurface.
The resulting model (model 1 in the following) remained stable while,
during a run corresponding to 120 ps, 500 configurations separated by
10 time steps of 100 a.u. (24 fs) were stored for subsequent
analysis.
In the last run, we used the SCC correction and increased the basis
set by including the d orbitals for sulphur atoms.
This basis set extension provided a major impovement in the
description of hypervalent sulphur molecules \cite{Niehaus_01} as well
as silicon-oxygen compounds \cite{Kaschner_97}.
While d orbitals on sulphur atoms give noticeable improvement, it
appears that they are less important for arsenic and we restrict
ourselves to including only s and p orbitals for the As atoms in order
to speed up the calculations.

Model 1 contains three topologically identical coordination defects,
namely intimate valence alternation pairs (IVAPs), where a singly
coordinated sulphur atom is attached to a three-fold coordinated
arsenic atom, thus increasing the coordination number of this arsenic
atom to four.
Apart from the IVAPs, the amorphous network is topologically ideal, in
the sense that each sulphur atom is bonded to two neighbors and each
arsenic atom is bonded to three.
There is, however, a certain degree of chemical disorder in this
system which manifests itself in the presence of nine As-As and six
S-S homopolar bonds.

In the context of photoinduced metastability, a great deal of
significance is attributed to the presence of topological and/or
chemical defects \cite{Kolobov_PIMAS}.
It is therefore imperative to create models both with and without such
defects in a theoretical investigation that attempts to be conclusive.
We produced additional models by ``surgically'' removing the IVAP
defects and homopolar bonds from model 1.
Model 2, which does not contain any topological defects, was obtained
by removing the three singly coordinated sulphur atoms from model 1
and rescaling this 197-atom model to the original density.
This procedure did not affect the stability of the amorphous network.

In order to eliminate the chemical defects, we iteratively applied
the following algorithm that utilizes the ideas
\cite{Mousseau_private} used to create models of binary amorphous
solids (e.g. a-SiO$_2$) from one-component continuous random networks
(e.g. a-Si) \cite{Barkema_2000}.
First, a sulphur atom was inserted in the middle of each As-As
homopolar bond.
Second, each S-S bond was replaced by a single sulphur atom located at
its mid-point so that each local 
As-S-S-As configuration turned into As-S-As.
Third, the distance between each newly introduced S atom and its two
nearest arsenic atoms in the newly created As-S-As units was reduced in
order to increase the bonding character of the As-S bonds stretched by
the above manipulation.
We set a constraint on the length of the modified As-S bonds so that
it did not exceed 2.5~\AA.
Fourth, the modified configuration was relaxed in an MD run at
$T=300$~K for 2.5~ps, until the potential energy reached a plateau.
After the first iteration, the 200-atom sample (that we call model 3
in the following) contained only one As-As and one S-S bond that were
spatially well separated (the minimum As-S distance among these four
atoms was 4.6~\AA).
Only two iterations were sufficient in order to obtain a model with
all-heteropolar bonds, which we refer to as model 4 in the following.
The defect statistics for models 1-4 are summarized in
Table~\ref{tab:models}.

\subsection{Data analysis}

A common way to assess the quality of a structural model is to compare
experimental and calculated static structure factors.
We have calculated the structure factor $S(Q)$ by Fourier transforming
the radial pair-correlation function $g(r)$ (also often called the
pair- or radial distribution function) defined as (see, e.g.,
Refs.~\onlinecite{Waseda_SNCM,Hansen_TSL,McQuarrie_SM}):
\begin{equation}
g(r) = \frac{V}{4\pi r^2 N^2} 
\left\langle 
\sum_{i\ne j} \frac{b_i b_j}{\langle b \rangle^2} \delta(r - r_{ij}) 
\right\rangle,
\label{eq:gofr} 
\end{equation}
where the sum is over all pairs of atoms in the sample of volume $V$
separated by distance $r_{ij}$, $N$ is the total number of atoms and
the angular brackets denote an ensemble average. In the case of
neutron scattering, that is of interest here, $b_i$ is the coherent
scattering length of atom $i$ and $\langle b \rangle$ is the average
scattering length.
This function gives the probability of finding a pair of atoms a
distance $r$ apart, relative to the probability expected for a
completely random distribution of atoms at the same density.
For a binary alloy, e.g. As$_2$S$_3$, it is of interest to decompose
$g(r)$ in terms of the partial pair-correlation functions
$g_{\alpha\beta}(r)$:
\begin{equation}
g(r) = \sum_{\alpha} \sum_{\beta} \bar{b}_{\alpha} \bar{b}_{\beta} g_{\alpha\beta}(r),
\end{equation}
where the double sum is over atomic types and $\bar{b}_{\alpha} =
c_{\alpha} b_{\alpha}/\langle b \rangle$, with $c_{\alpha} =
N_{\alpha}/N$ being the atomic fraction of $\alpha$ atoms.
From Eq.~(\ref{eq:gofr}), it follows that
\begin{equation}
g_{\alpha\beta}(r) = \frac{V}{4\pi r^2 N^2 c_{\alpha} c_{\beta}} 
\left\langle 
\sum_{i{\alpha} \ne j{\beta}} \delta(r - r_{i{\alpha},j{\beta}}) 
\right\rangle,
\label{eq:gabofr}
\end{equation}
where the index $i\alpha$ runs over $\alpha$-type atoms only.
The values of the scattering lengths used here were $b_{\mathrm{As}}=6.58$~fm
and $b_{\mathrm{S}} = 2.847$~fm (see, e.g., Ref.~\onlinecite{nistscatt}).
In practice, we use a standard algorithm, where the $\delta$-function in
Eq.~(\ref{eq:gabofr}) is replaced by a function which is non-zero in a
small range of separations, and a histogram is compiled of all pair
separations falling within each such range (see
e.g. Ref.~\onlinecite{Allen_CSL}).
Analogously, the bond-angle distribution function can be calculated as
a histogram of all bond angles in the system.

While bond-angle distributions provide information on the short-range
order of an amorphous material, ring statistics have been generally
used as a measure of the medium-range order.
An $n$-membered ring is a closed loop with $n$ atoms (or bonds).
Here, we count only the shortest-path (irreducible) rings
\cite{Franzblau_91}, i.e.  those which do not have ``shortcuts''
across them.
In order to identify such rings we use the algorithm due to Franzblau
\cite{Franzblau_91}, as implemented in the program ``statix'' by
Jungnickel \cite{Jungnickel_statix,Jungnickel_94}.
The basic idea is to travel along the network paths (bond chains)
containing a tagged atom and identify all the rings of length up to a
given maximum.
For each ring, it can then be verified whether it is an irreducible
one (containing no cross links).

Experiments also provide information on the vibrational and electronic
densities of states (VDOS and EDOS) which can be compared with the
results obtained from simulation.
In addition to this, simulations allow one to assess the degree of
localization of the vibrational and electronic eigenstates.
We compute both the vibrational and electronic densities of states 
by using the following definition:
\begin{equation}
g(\omega) = C \sum_n \delta(\omega - \omega_{n}),
\label{eq:dos}
\end{equation}
where the constant $C$ is determined by normalization, $\omega_n$ are
eigenfrequencies of the Hessian (dynamical) matrix of an
energy-minimum configuration in the case of the VDOS or Kohn-Sham
eigenfrequencies (or energies) corresponding to this configuration in
the case of the EDOS, and the sum is over all eigenstates.
In practice, in order to obtain a smooth representaion of $g(\omega)$,
the delta function in Eq.~(\ref{eq:dos}) is replaced by a Gaussian
function centered at $\omega_n$.

In contrast to the eigenstates of a perfect crystal, that extend over
the entire sample, some eigenstates in disordered solids are localized
at relatively small groups of atoms.
The degree of localization can be quantified by the inverse
participation ratio that is defined in terms of dynamical-matrix
eigenmodes or Mulliken partial charges, for vibrational or electronic
excitations respectively.
For a vibrational mode $n$, the inverse participation ratio can be
defined \cite{Bell_72} as
\begin{equation}
p^{-1}_n = 
\left. 
\left(\sum_{i=1}^{N} |\mathbf{u}_i^{(n)}|^4 \right)
\right/ 
\left(\sum_{i=1}^{N} |\mathbf{u}_i^{(n)}|^2 \right)^2.
\end{equation}
When the displacement eigenvectors $\mathbf{u}_i^{(n)}$, $n=1,2, \dots,3N$,
are normalized to unity ($\sum_i|\mathbf{u}_i^{(n)}|^2 = 1$), $p^{-1}_n =
1$ for a mode totally localized at one atom and $p^{-1}_n = 1/N$ for a
completely extended mode, such as a rigid-body displacement.

In the case of the electronic properties, the linear combination of
atomic orbitals (LCAO) concept employed in the DFTB program allows one
to separate the contributions from individual atomic sites and
orbitals to the total charge for a particular eigenstate, and to
decompose the total EDOS in terms of the local electronic densities of
states (LEDOS).
Using the Mulliken population analysis (see, e.g.,
Ref.~\onlinecite{Szabo_MQC}), the inverse participation ratio for an
electronic state $n$ can be written as \cite{Cobb_96}
\begin{equation}
p^{-1}_n = 
\sum_{i=1}^N |q_i^{(n)}|^2,
\label{eq:eipr}
\end{equation}
where the contribution to state $n$ from atomic site $i$, $q_i^{(n)}$,
can be expressed in terms of the wavefunction coefficients in the
tight-binding basis $c_{\mu n}$ and the elements of the
overlap matrix $\mathbf{S}$ :
\begin{equation}
q_i^{(n)} = \sum_{\mu\in i, \nu} 
S_{\mu\nu}c_{\mu n}c_{\nu n}
\label{eq:partcharges}
\end{equation}
Here, in the double sum, the index $\mu$ runs only over the
atomic orbitals located on atom $i$ and the index $\nu$ goes over all orbitals.
As in the case of the vibrational inverse participation ratio, the
electronic $p^{-1}_n$ is equal to $1/N$ for a totally delocalized mode
and approaches unity with increasing degree of localization.
The partial charges $q_i^{(n)}$ allow one to detect on which atoms
most of the charge is localized for a particular eigenstate.
By refraining from summing over the atomic orbitals, i.e.  over
$\mu$ in Eq.~(\ref{eq:partcharges}), one can identify the type of
the atomic orbitals, e.g. s or p, most actively participating in an
eigenstate.
Analogously, the local EDOS for a particular orbital type can 
be obtained via the following expression :
\begin{equation}
g_{\mu}(\omega) = 
C \sum_n \delta(\omega - \omega_{n}) \sum_{\nu} 
S_{\mu\nu}c_{\mu n}c_{\nu n}
\label{eq:ledos}
\end{equation}

\section{Results}

\subsection{Structure}

\begin{table}[t]
\begin{tabular}{p{4cm}*{4}{r}}
\\ \hline \hline 
\hfill {Model} &   1 &   2 &   3 &   4 \\
\hline 
No. of atoms  & 200 & 197 & 200 & 200 \\
No. of As-As 
bonds         &   9 &   9 &   1 &   0 \\
No. of S-S
bonds         &   6 &   6 &   1 &   0 \\
No. of IVAPs  &   3 &   0 &   0 &   0 \\
Total No. of
defects       &  18 &  15 &   2 &   0 \\
\hline \hline
\end{tabular}
\caption{Defect statistics for models 1-4.}
\label{tab:models}
\end{table}

\begin{figure}[t] 
\centerline{(a)\includegraphics[width=7.5cm]{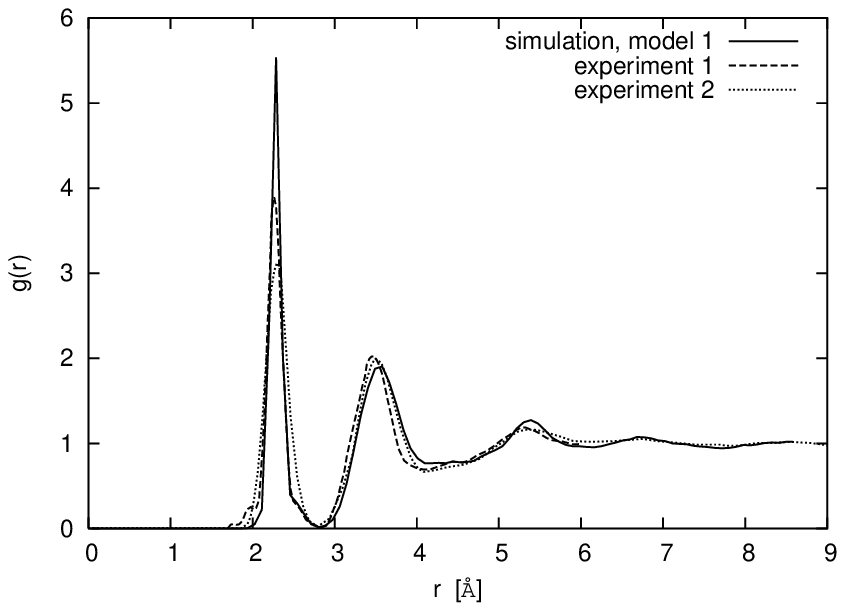}}
\centerline{(b)\includegraphics[width=7.5cm]{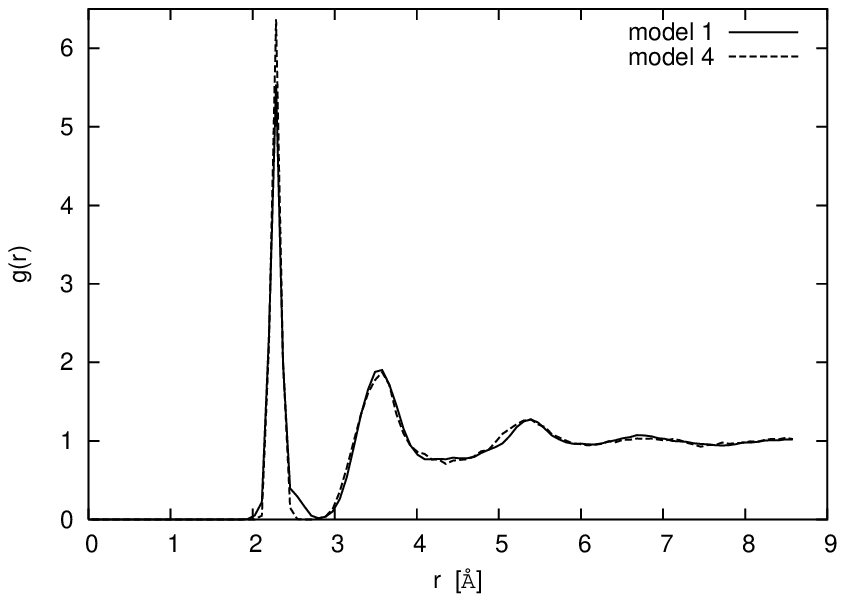}}
\centerline{(c)\includegraphics[width=7.5cm]{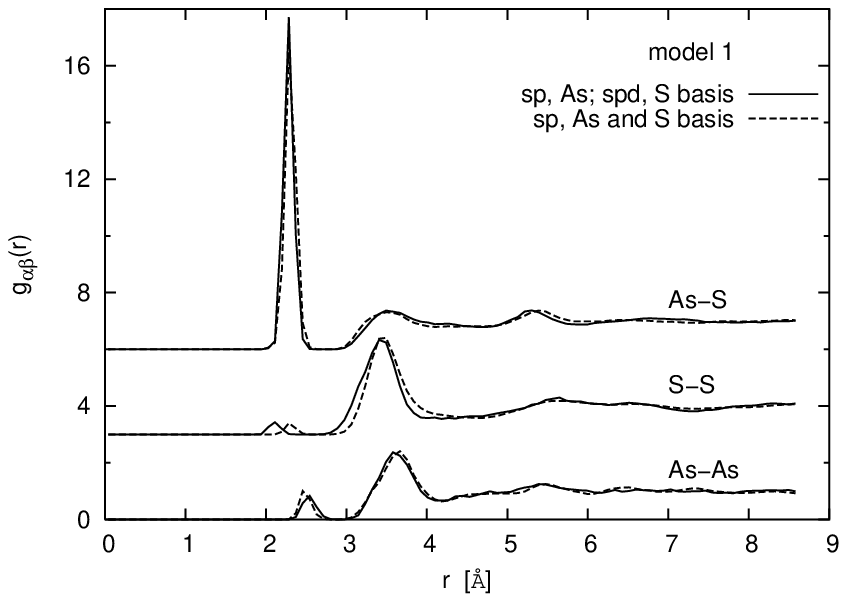}}
\caption{(a) Total pair-correlation functions for model 1 and the
neutron-diffraction experiments 1 (Ref.~\onlinecite{Lee_98}) and 2
(Ref.~\onlinecite{Iwadate_99}).  (b) Total pair-correlation functions
for models 1 (the same as in (a)) and 4.  (c) Partial pair-correlation
functions for model 1.  $g_{\mathrm{S-S}}(r)$ and
$g_{\mathrm{As-S}}(r)$ are shifted upwards by 3 and 6 units,
respectively.}
\label{fig:pdf200d}
\end{figure}

\begin{figure}[t] 
\centerline{(a)\includegraphics[width=7.5cm]{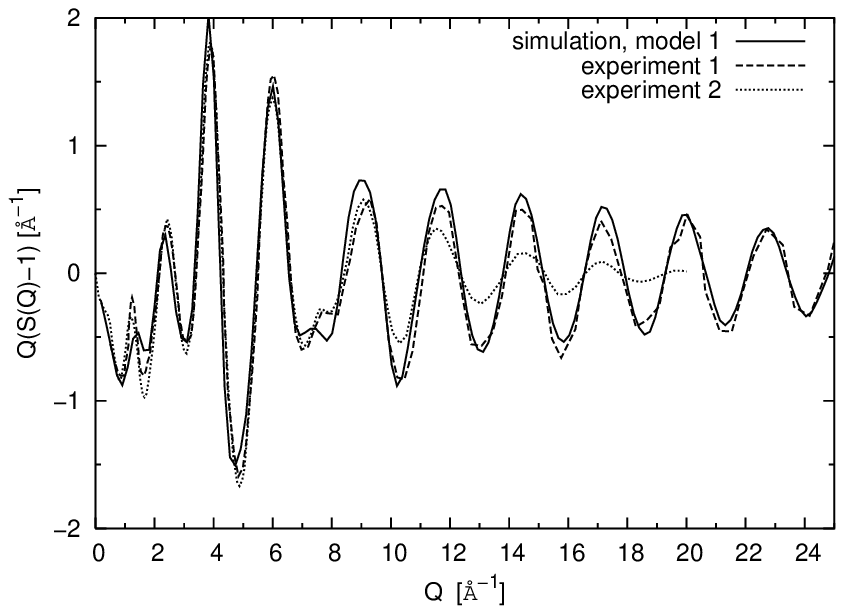}}
\centerline{(b)\includegraphics[width=7.5cm]{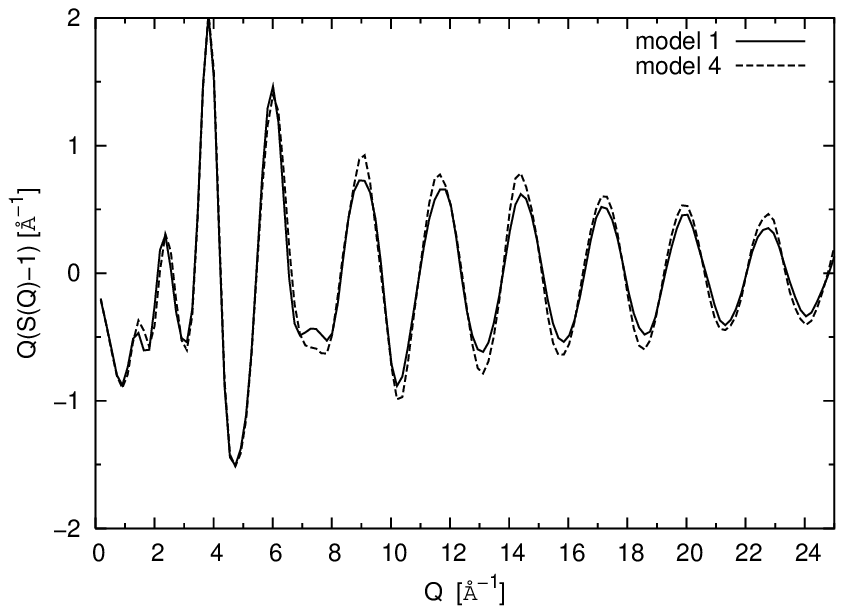}}
\centerline{(c)\includegraphics[width=7.5cm]{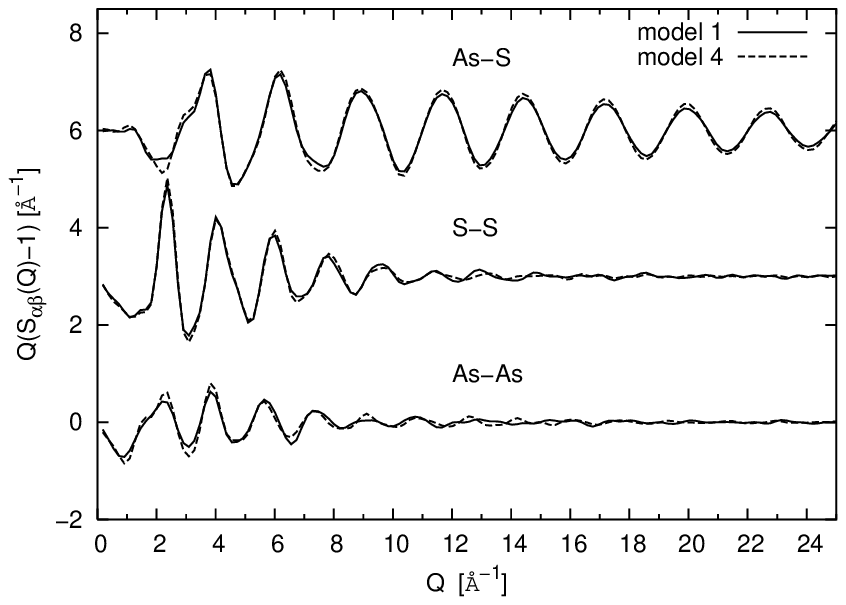}}
\caption{(a) Reduced structure factors (interference functions) for
model 1 and the neutron-diffraction experiments 1
(Ref.~\onlinecite{Lee_98}) and 2 (Ref.~\onlinecite{Iwadate_99}).  (b)
Total interference functions for models 1 (the same as in (a)) and 4.
(c) Partial interference functions for models 1 and 4.  The functions
corresponding to S-S and As-S correlations are shifted upwards by 3
and 6 units, respectively.}
\label{fig:f200d}
\end{figure}

\begin{figure*}[t] 
\centerline{(a)\includegraphics[width=7.5cm]{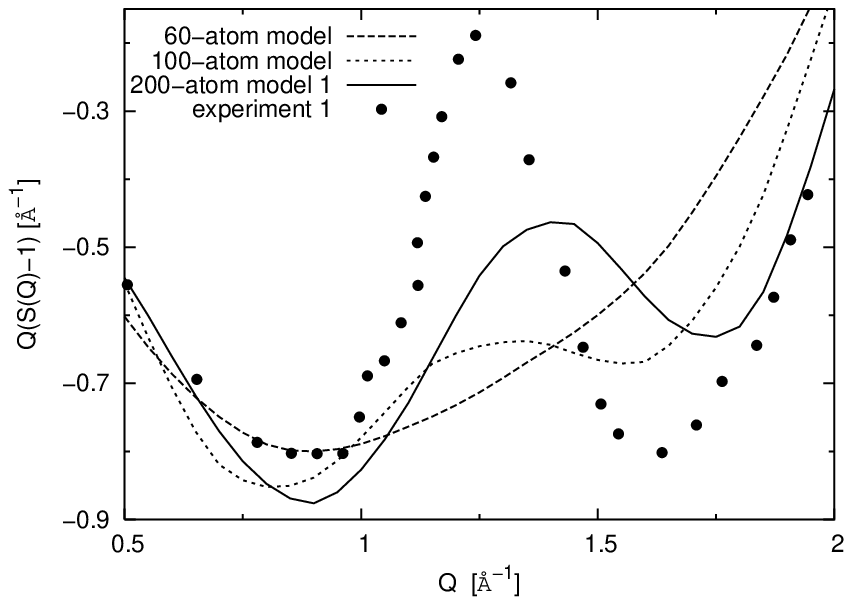} \hspace{0.5cm}
            (c)\includegraphics[width=7.5cm]{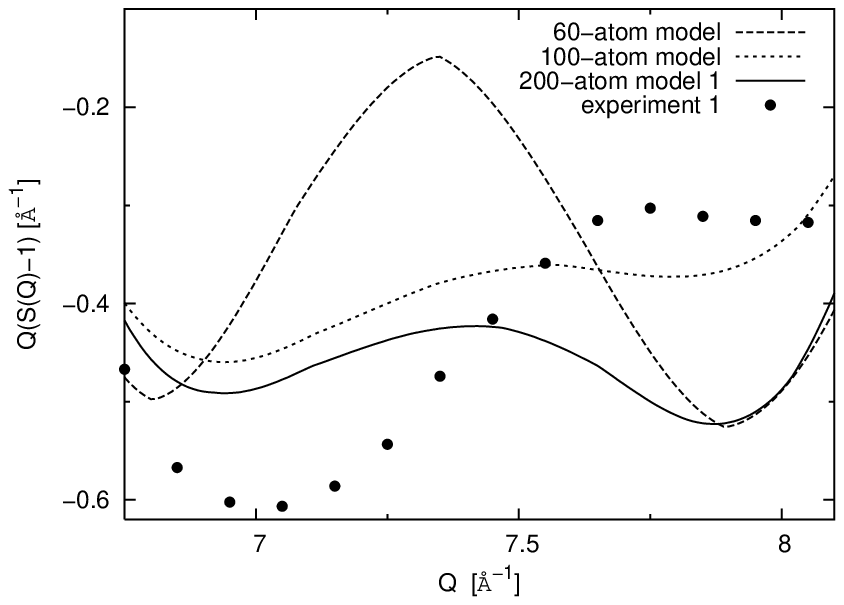}}
\centerline{(b)\includegraphics[width=7.5cm]{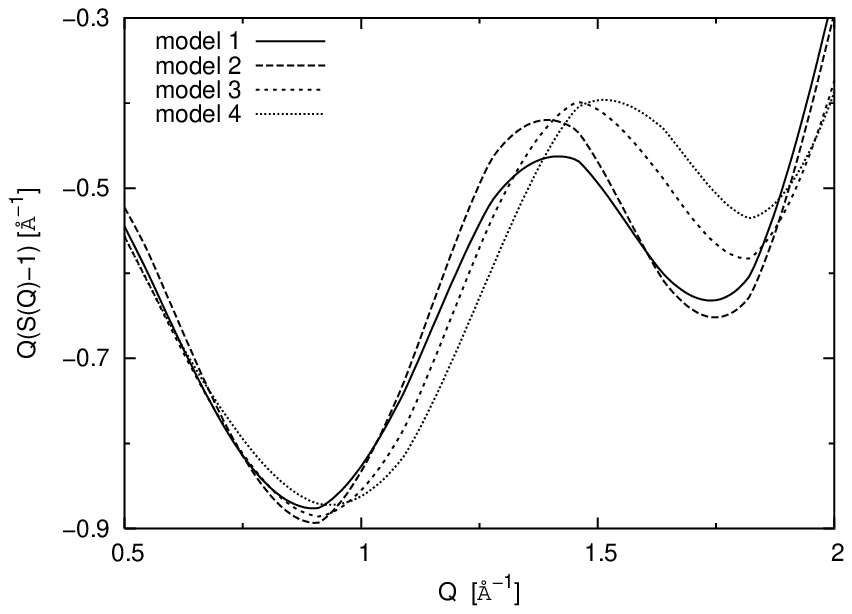} \hspace{0.5cm}
            (d)\includegraphics[width=7.5cm]{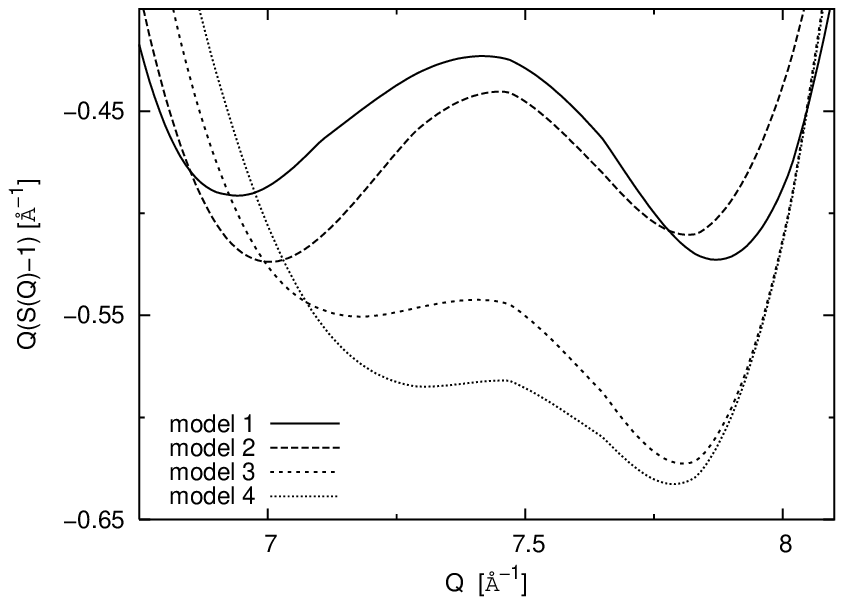}}
\caption{Close-up of the parts of $F(Q)=Q(S(Q)-1)$ most sensitive to
changes in the local structure. The FSDP region for (a) the 60- and
100-atom models, model 1 and experiment 1, and (b) models 1-4,
respectively. The region near $Q \approx 7.5$~\AA$^{-1}$ for (c) the
60- and 100-atom models, model 1 and experiment 1, and (d) models 1-4,
respectively.  }
\label{fig:fsdp}
\end{figure*}

By using the construction method described in Sec.~\ref{sec:prep}, we
obtained four models of a-As$_2$S$_3$ which are distinguished by the
presence and concentration of topological and chemical defects (see
Table~\ref{tab:models}).
We also created 60- and 100-atom models with similar concentrations of
homopolar bonds as in models 1 and 2, and without coordination
defects.

Fig.~\ref{fig:pdf200d}(a) shows that the pair-correlation function
(PCF) corresponding to model 1 compares well with two independent
neutron-scattering experimental results.
The discrepancy between the two experimental PCFs allows one to
estimate the uncertainty in the experimental data.
The main difference between the experimental and simulated 
results is in the height of the first peak.
Since the experimental PCFs are obtained by Fourier transforming the
measured static structure factor, where the large-$Q$ oscillations are
damped by applying a window function, this reduces the height of the
first peak in $g(r)$ and also broadens its width.

\begin{table}[t]
\begin{tabular}{p{2.5cm}*{3}{p{1.5cm}}}
\\ \hline \hline 
Sample        & $r_{\mathrm{As-S}}$ [\AA] & $Z_{\mathrm{As-S}}$ & $Z_{\mathrm{S-As}}$ \\
\hline 
experiment 1  & 2.27  & 2.8  & 1.8 \\
model 1       & 2.279 & 2.81 & 1.88 \\
\hline \hline
\end{tabular}
\caption{Values of average bond length and coordination numbers for
the first coordination shell found from peak fitting (experiment 1)
and direct calculation (model 1).}
\label{tab:ZAsS}
\end{table}

Although the PCFs corresponding to models 2-4 are quite similar to
that for model 1 (which is why we do not show here the PCFs for models
2 and 3), there is one conspicuous distinction in the shape of the
first peak that depends on whether or not the system contains
homopolar bonds (see Fig.~\ref{fig:pdf200d}(b)).
While this peak is symmetric for model 4 with all heteropolar bonds,
there is a shoulder on either side of the peak in the PCF for model 1.
From Fig.~\ref{fig:pdf200d}(c), it is seen that the shoulders in the
first peak of the total PCF originate from the homopolar As-As and S-S
bonds which produce small peaks in the respective partial PCFs 
at this position.
It is remarkable that the calculated PCF for model 1 virtually
reproduces the right-hand side of the first peak in the PCF from
experiment 1 (see Fig.~\ref{fig:pdf200d}(a)).
This result supports one of the conclusions of
Ref.~\onlinecite{Lee_98} (experiment 1) that the low- and high-$r$
sides of the base of the first peak in the PCF can be ascribed to S-S
and As-As bonds, respectively.
Table~\ref{tab:ZAsS} further demonstrates that the agreement between
the structural characteristics of model 1 and experiment 1 is very good
and, in particular, that the As-S coordination numbers for these two
samples are practically the same.
The larger discrepancy in the S-As coordination number and the low-$r$
side of the first peak in the PCF arises from the relatively small
system size and the fact that the numbers of As-As (nine) and S-S
(six) homopolar bonds are not equal to each other in model~1.
The position of the first peak in the partial PCF $g_{\mathrm{S-S}}(r)$
also plays a role here.
As is seen from Fig.~\ref{fig:pdf200d}(c), this peak shifts towards
the low-$r$ end when d orbitals on sulphur atoms are included into the
basis set.
When a $T=300$~K run is performed without the d orbitals in the basis set,
the position of the first peak in $g_{\mathrm{S-S}}$ coincides with
that of the first peak in $g_{\mathrm{As-S}}$, and the shoulder on the
low-$r$ side of the simulated total $g(r)$ is not seen at all.

\begin{figure*}[t] 
\centerline{\includegraphics[width=16cm]{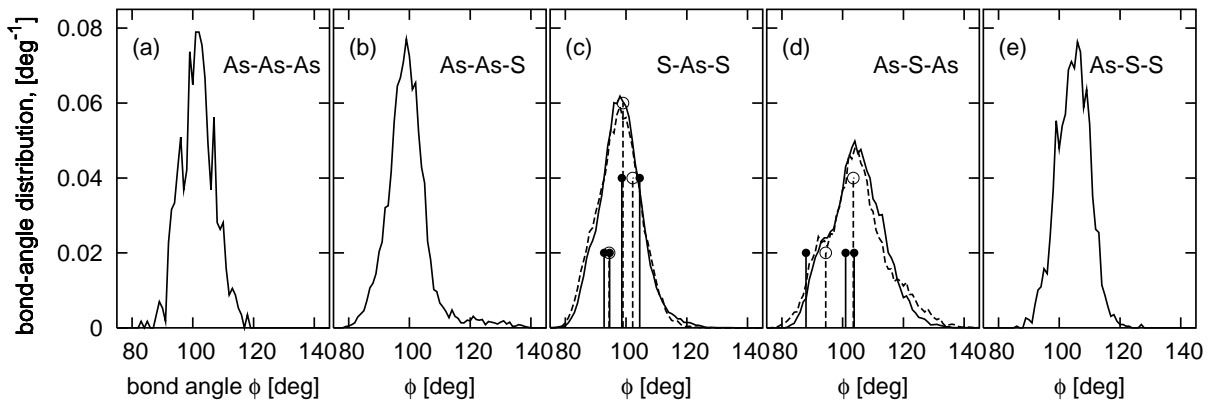}}
\caption{Bond-angle distributions. Solid lines in (a)-(e) and dashed
lines in (c) and (d) correspond to models 1 and 4 respectively.  The
height of vertical lines topped by circles is proportional to the
number of distinct angles within one degree in the crystal structure
of orpiment. Solid vertical lines and circles correspond to the
experimental data from the crystallographic database and
Ref.~\onlinecite{Mullen_72}, and the dashed vertical lines and open
circles correspond to the structure optimized by the DFTB method.}
\label{fig:bad}
\end{figure*}

It is instructive to compare the structural data also in $Q$-space, as
this often emphasizes features that are not obvious in an $r$-space
representation.
Fig.~\ref{fig:f200d}(a) shows the reduced structure factors (or
interference functions) $F(Q) = Q(S(Q)-1)$, related to the PCFs in
Fig.~\ref{fig:pdf200d}(a) by a Fourier transform.
Again, the agreement between model 1 and experiment 1 is very 
good.
From the rate of decay of $F(Q)$ from experiment 2, it is apparent
that the data in Ref.~\onlinecite{Iwadate_99} are reported for $F(Q)$
multiplied by a window function.

$F(Q)$ for model~1 exhibits a first-sharp diffraction peak (FSDP) at
about the same position, $Q \approx 1.5$~\AA$^{-1}$, as found in the
experiments.
The magnitude of this peak depends on the system size and is expected
to increase for a larger model.
This statement is supported by the observation of this tendency as the
peak develops in our 60-, 100-, and 200-atom models (see
Fig.~\ref{fig:fsdp}(a)).
Fig.~\ref{fig:fsdp}(b) demonstrates a systematic displacement in the
position of the FSDP towards the high-$Q$ end as the number of defects
is reduced from model 1 to model~4.
In Ref.~\onlinecite{Tanaka_75}, it was observed that the FSDP in X-ray
diffraction intensity curves for an a-As$_2$S$_3$ thick (8~$\mu$m) film
reversibly moved towards the low- and high-$Q$ end upon illumination
and annealing, respectively.
The result in Fig.~\ref{fig:fsdp}(b) is consistent with the above
observation, if we suppose that the defect concentration is reversibly
increased and decreased upon illumination and annealing, respectively.
The height of the measured \cite{Tanaka_75} peak, however, increased
as its position wavenumber, $Q_1$, decreased after illumination, while
in the simulations we observed the opposite tendency for the height of
the FSDP to increase upon elimination of defects and increasing
$Q_1$.
A possible reason for the different behavior of the height of the FSDP
is that the simulations of the bulk a-As$_2$S$_3$ were performed at
constant volume, while the experiment was done for an amorphous film
at atmospheric pressure.

Another distinction between the different interference functions
presented in Fig.~\ref{fig:f200d} is seen in the range 
7~\AA$^{-1}$~$\lesssim$~$Q$~$\lesssim$~8~\AA$^{-1}$.
$F(Q)$ appears to be very sensitive to structural differences in this
particular range of wavenumbers, as is apparent from
Fig.~\ref{fig:f200d}(b), where the interference function for model 1
is compared with that for model 4, and from Fig.~\ref{fig:fsdp}(c,d),
where $F(Q)$ is magnified in this $Q$-interval, and the differences
between the curves are prominent.
Although the differences between the partial interference functions
for models 1 and 4 (shown in Fig.~\ref{fig:f200d}(c)) are each rather
subtle in this $Q$-interval, they become more pronounced when combined
into the total $F(Q)$ (see Fig.~\ref{fig:f200d}(b)).
The small peak, which is seen in $F(Q)$ for model 1 at $Q \approx
7.5$~\AA$^{-1}$, diminishes from model 1 to 4, so that it is
practically not seen in the case of the stoichiometric model 4 (see
Fig.~\ref{fig:fsdp}(d)).
The presence and position of this small peak may be attributed to the
presence and spatial distribution, respectively, of homopolar bonds in
the system.
Similar peaks are seen in $F(Q)$ at about $Q=7.5$~\AA$^{-1}$ for both
experiments mentioned here (Fig.~\ref{fig:f200d}(a),
Refs.~\onlinecite{Lee_98,Iwadate_99}) and in the experiment reported
in Ref.~\onlinecite{Haar_2000}.
While, in all these independent experiments, these peaks virtually
coincide, in the different models presented here they do not agree so
well (see Fig.~\ref{fig:fsdp}(c)), as, perhaps, can be expected in
the case of small system sizes.

Another structural characteristic that is of interest, and that is
easily accessible in computer simulations, is the bond-angle
distribution (see Fig.~\ref{fig:bad}).
Although we have no experimental data for this distribution, the known
structure of a corresponding crystal can serve as a guide in assessing
the quality of our models~--- many statistical distributions
associated with amorphous models agree overall with the respective
broadened distributions for the counterpart crystals (see, e.g.,
Refs.~\onlinecite{Simdyankin_PRB_2002,Simdyankin_00}), which also
applies to the bond-angle distributions presented here.
In addition to this, Fig.~\ref{fig:bad}(c),(d) shows that the geometry
optimization with DFTB results in a crystal structure that agrees with
the experimental one to within about two percent, not only in bond
distances and lattice constants, but also in bond angles.

It may appear reasonable to ascribe the asymmetry of the distributions
for S-As-S and As-S-As angles in the form of the shoulders on the
low-angle side to the presence of four-membered As-S-As-S rings, as
has been observed in tetrahedrally bonded chalcogenide semiconductors,
e.g. GeS$_2$ \cite{Blaineau_2003}.
Although such four-membered rings would definitely contribute to the
low-angle part of the distribution due to geometrical constraints,
their number in our models is not so large (see Table~\ref{tab:rings})
and, for that reason, the relative fraction of angles involved in
these rings is rather small (about 7~\% of S-As-S angles and 13~\% of
As-S-As angles).
We therefore conclude that the asymmetry of the heteropolar bond-angle
distributions is inherent to this type of material, as is evidenced in
the case of both crystal and amorphous structures.

\begin{table}[t]
\begin{tabular}{p{2cm}*{13}{r}}
\\ \hline \hline 
ring size $n$ & 4 & 5 & 6 & 8 &10 &14 &15 &16 &18 &19 &20 &21 &22 \\
\hline 
models 1, 2 & 5 & 6 & 1 & 1 & 1 & 2 & 2 & 3 & 1 & 2 & 6 & 1 & 1 \\
model 3     & 8 & 0 & 5 & 1 & 1 & 2 & 2 & 3 & 4 & 0 & 6 & 0 & 3 \\
model 4     & 8 & 0 & 5 & 1 & 1 & 2 & 0 & 5 & 4 & 0 & 6 & 0 & 3 \\
\hline \hline
\end{tabular}
\caption{Ring statistics - number of $n$-membered shortest path
\cite{Franzblau_91} rings for $n \le 22$. Columns containing only
zeros are not included.}
\label{tab:rings}
\end{table}

Interestingly, there are no 12-membered rings in our models of
a-As$_2$S$_3$, while only such rings exist in the crystal structure of
orpiment.
This is another strong piece of evidence that models 1-4 do not
contain any memory of the initial crystalline atomic arrangement.
Also there are no three-membered rings, that were previously reported
to be found in a model of a-As$_2$Se$_3$ \cite{Drabold_PRB_00}, whose
presence would contribute a few small angles and would increase the
number of homopolar bonds.

\begin{figure}[t] 
\centerline{\includegraphics[width=6cm]{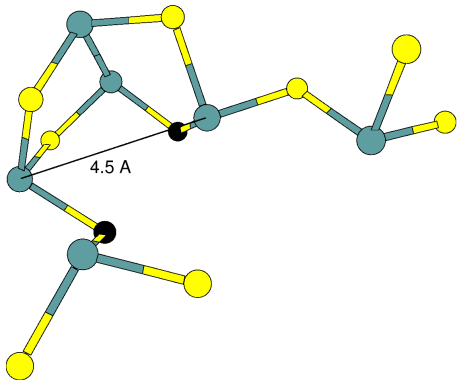}}
\caption{Fragment of model 1: two bond-sharing five-membered rings and
the two AsS$_3$ groups connected to this structure.  The shading of
the As atoms (all with three neighbors) is darker than that of the S
atoms (all with two neighbors).  The HOMO level is mostly localized on
the two S atoms that are marked black.  The distance between these two
atoms is 3.42 \AA.}
\label{fig:two5rings}
\end{figure}

Special significance can be attributed to the presence of
five-membered rings in models 1 and 2 with an appreciable
concentration of homopolar bonds (models with all-heteropolar bonds
contain only an even number of atoms in all rings).
When such rings share some of the bonds, the resulting local structure
is close to that of cage-like molecules (e.g. As$_4$S$_4$ or
As$_4$S$_3$), as found in the vapor phase and in some chalcogenide
molecular crystals.
Fig~\ref{fig:two5rings} (cf. Fig.~8.8(b) in
Ref.~\onlinecite{Yannopoulos_2003}) shows two such bond-sharing rings.
Upon breaking the two bonds connecting the rings to the rest of the
network, the distance of 4.54~\AA ~between the two freed arsenic atoms
could be reduced, thus producing another As-As homopolar bond and this
group of atoms would then form an As$_4$S$_4$ molecule.
Evidence of the presence of such molecules in bulk
As$_{\mathbf{x}}$S$_{1-\mathbf{x}}$ glasses from Raman-scattering
experiments has recently been reported in
Ref.~\onlinecite{Georgiev_2003}.
Our result shows that the As$_4$S$_4$ fragments may not only form
discrete cage-like molecules but also be embedded into the amorphous network.
We verified that the vibrational signatures of the As$_4$S$_4$
fragment from models 1 and 2 are similar to those from an isolated
As$_4$S$_4$ molecule, apart from a few very symmetric modes of the
latter.

\subsection{Vibrational properties}
\label{sec:vibr}

\begin{figure}[t] 
\centerline{\includegraphics[width=7.5cm]{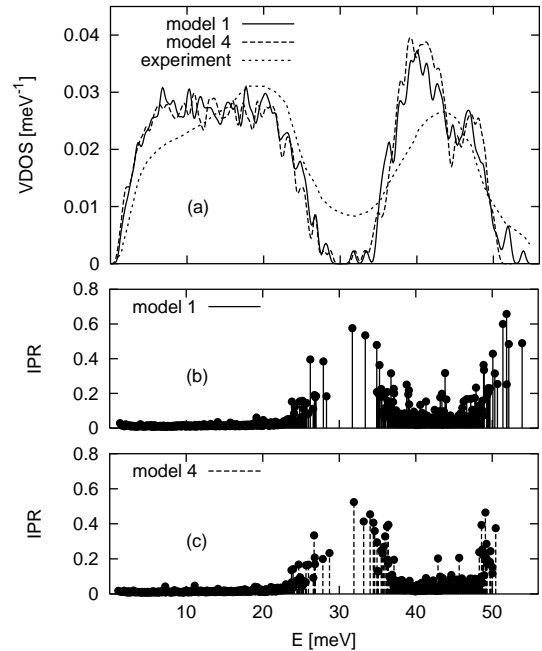}}
\caption{Vibrational densities of states (a) and inverse participation
ratios (b),(c) for models 1 and 4. The experimental data in (a) are
obtained from Ref.~\onlinecite{Isakov_93}.}
\label{fig:vdos}
\end{figure}

The vibrational density of states (VDOS) for models~1 and 4 is shown
in Fig.~\ref{fig:vdos}(a).
It has the two-band form generally observed in amorphous
semiconductors.
The vibrational spectrum for our models is essentially superimposable
on the calculated VDOS \cite{Drabold_PRB_00} for a model of
a-As$_2$Se$_3$ if the energy in Fig.~\ref{fig:vdos}(a) is downscaled
by a factor of about 0.67.
All main features~--- the position of the gap between the acoustic and
optic bands, as well as the relative width and height of VDOS within
these bands~--- agree with available inelastic neutron-scattering
experimental data \cite{Isakov_93,Haar_2000}.
Note that the experimental curve in Fig.~\ref{fig:vdos}(a) corresponds
to a measurement at room temperature, whereas the results obtained
from simulation are calculated for an energy-minimum configuration in
the harmonic approximation.
An attempt to measure the VDOS of a-As$_2$S$_3$ at temperatures as low
as 25~K was made in Ref.~\onlinecite{Haar_2000}, and the tendency for
the narrowing and heightening of the optic band and the flattening of
the top of the acoustic band was captured, although the experimental
uncertainty was rather large.

The inverse participation ratios (IPR) (see
Fig.~\ref{fig:vdos}(b),(c)) show that the vibrational eigenmodes are
significantly localized in the gap between the acoustic and optic
bands, and the high-$E$ end of the spectrum.
The first three highest energy modes in model~1 are localized on S-S
bonds, while the next two (with the largest IPR) are localized on
IVAPs.
The VDOS's for models 1 and 4 differ mainly in the absence of the
just-mentioned highest energy modes from the spectrum corresponding to
the stoichiometric model 4.
The highest energy mode ($E=50.5$~meV) in model 4 is localized on a
relatively complex structure involving three AsS$_3$ pyramids in a
chain As-S-As-S-As, where both As-S-As angles (118 and 122$^\circ$)
are at the large-$\phi$ end of the bond-angle distribution shown in
Fig.~\ref{fig:bad}(d).
Within the optic band, the modes with IPR greater than 0.2, at
$E=43.8$~meV in model~1 and at $E=42.9$ and 45.7~meV in model~4, are
localized at four-membered rings.
In the band gap, the mode at the top of the acoustic band in model~1
is localized on an As-As-S-S chain, and the mode at the bottom of the
optic band is localized on an IVAP.
The mode at the top of the acoustic band in model~4 is predominantly
localized on a four-membered ring, and the mode at the bottom of the
optic band is almost entirely localized on a six-membered ring.

\subsection{Electronic structure}

\begin{figure}[t] 
\centerline{(a)\includegraphics[width=8cm]{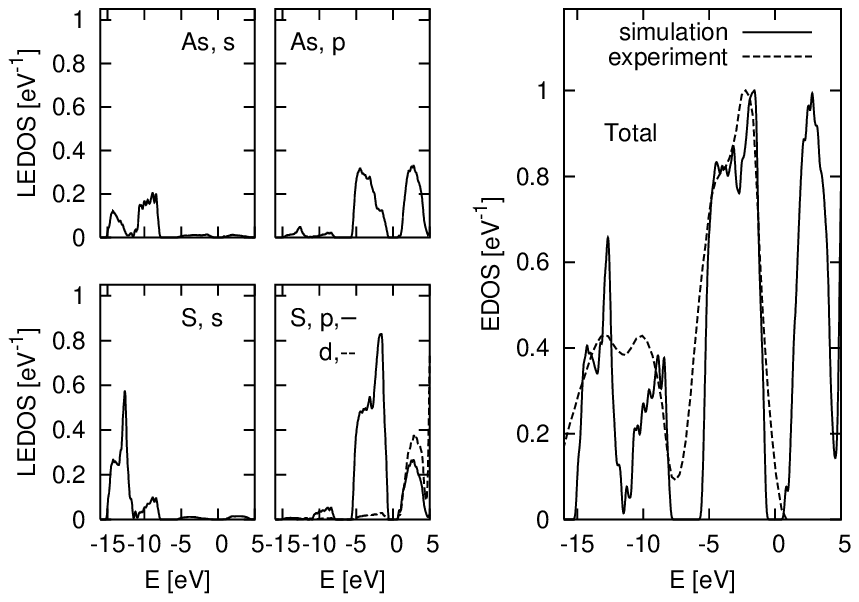}}
\centerline{(b)\includegraphics[width=8cm]{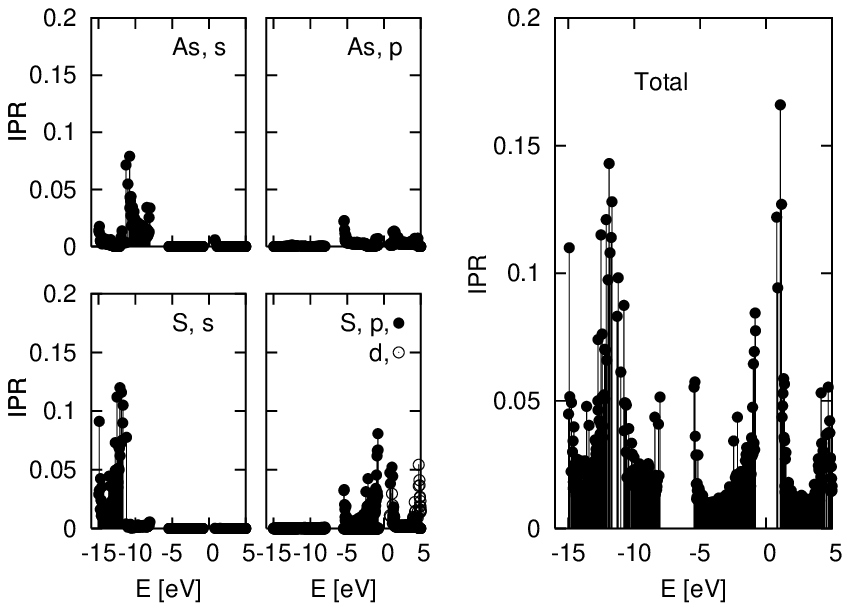}}
\caption{Local and total electronic density of states (a) and inverse
participation ratios (b) for model~1. The Fermi energy is at the
origin. The experimental data in (a) are
obtained from Ref.~\onlinecite{Bishop_75}.} 
\label{fig:ledos200d}
\end{figure}

\begin{figure}[t] 
\centerline{\includegraphics[width=7.5cm]{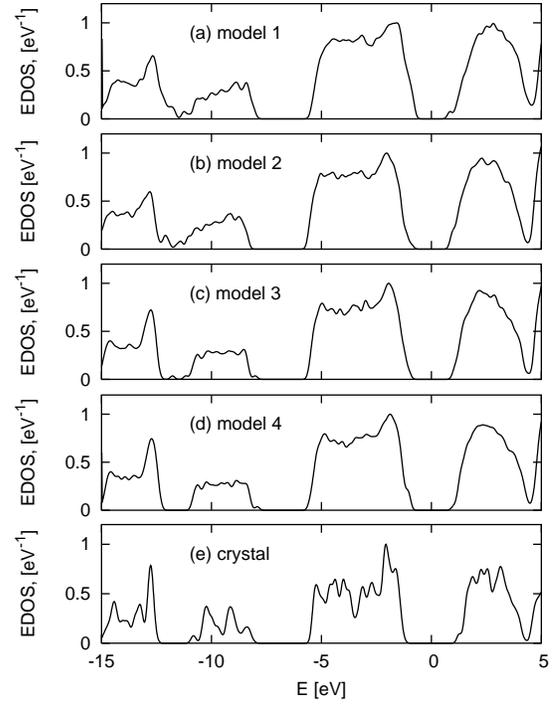}}
\caption{Electronic densities of states of the models of a-As$_2$S$_3$
and orpiment.}
\label{fig:edoses}
\end{figure}

\begin{figure}[t] 
\centerline{\includegraphics[width=7.5cm]{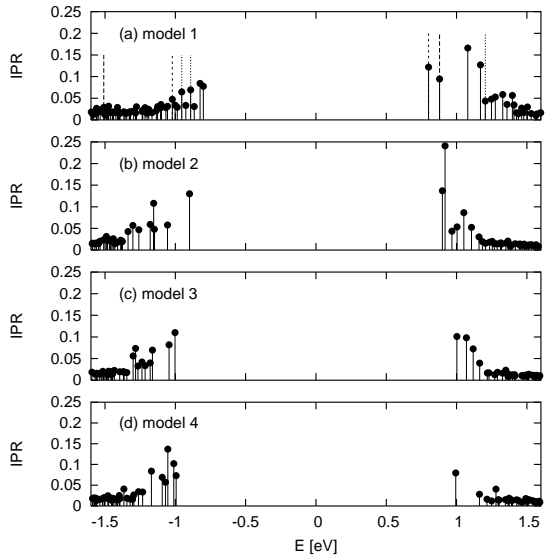}}
\caption{Inverse participation ratios at the band edges for all the
four models. The vertical dashed lines in (a) demarcate the levels 
mostly localized at IVAPs, with each line type corresponding to 
the same IVAP.}
\label{fig:eprs}
\end{figure}

The electronic density of states for model 1, as well as the inverse
participation ratios for these states, are shown in
Fig.~\ref{fig:ledos200d}.
The calculated total EDOS is in good agreement with the density of
valence states \cite{Bishop_75} measured by X-ray photoemission
spectroscopy.
The total EDOS is also very similar to that of arsenic selenide in
Ref.~\onlinecite{Drabold_PRB_00}, where an experimental result
\cite{Liang_75} for that material was presented to be in agreement
with the calculated one.

The local EDOS's for different elements and orbital types, shown in
Fig.~\ref{fig:ledos200d}(a), confirm the analysis presented in
Ref.~\onlinecite{Bishop_75}.
The top of the valence band is due to the non-bonding lone-pair p
orbitals of the S atoms, and the rest of the valence band is
attributed to the bonding p orbitals on the S and As atoms.

The s band is composed of two sub-bands.
The low-$E$ sub-band at about -(15-12)~eV is essentially an s type
sulphur band, and the sub-band at about -(12-8)~eV is predominately
due to the arsenic s orbitals.
In Ref.~\onlinecite{Drabold_PRB_00}, the fact that the s band of
selenium is below the s band of arsenic was attributed to the greater
nuclear charge of Se.
Since the electronic structure of arsenic sulphide, a compound
containing a much lighter chalcogen, virtually coincides with that for
arsenic selenide, the above explanation is incorrect.
A strong repulsion between As and S/Se s levels due to chemical
ordering is more likely to be responsible for the separation of the s
band into high (As) and low (S/Se) sub-bands \cite{Bishop_75}.
Note that the two s sub-bands are perfectly separated only when the
chemical ordering is perfect, i.e. in models with all-heteropolar
bonds, as seen in Fig.~\ref{fig:edoses}(d,e).
The degree of admixture of As and S s orbitals within the low and high
sub-bands, respectively, is about 30~\% (see the ``As, s'' and ``S,
s'' panels in Fig.~\ref{fig:ledos200d}(a)), again in agreement with
Ref.~\onlinecite{Bishop_75}.

The conduction band is composed of about equal contributions from the
antibonding As and S p orbitals and the S d orbitals.
Perhaps, if the As d orbitals were included in the basis set, there
would be a contribution from them too.
While this might not significantly affect the structural and
ground-state electronic properties of our models, this remark could be
of greater importance for excited-state simulations.

The EDOS's for models 1-4 and for the crystal structure of orpiment
are plotted in Fig.~\ref{fig:edoses}.
It is seen that the overall similarity is preserved for all these
structures.
The main differences among the EDOS's corresponding to the amorphous
structures are the widening of the optical band gap (see also
Table~\ref{tab:bandgaps}) and the clearing of the gap within the
low-energy s-band as the number of defects diminishes.

\begin{table}
\begin{tabular}{p{4cm}*{13}{r}}
\\ \hline \hline 
model               & 1    & 2    & 3    & 4    & crystal \\
\hline 
$E$(LUMO)-$E$(HOMO) & 1.60 & 1.80 & 2.00 & 1.99 & 2.51 \\
half-maximum gap    & 2.67 & 2.72 & 2.74 & 2.74 & 3.13 \\
\hline \hline
\end{tabular}
\caption{Band-gap energies in eV estimated as the differences between
the LUMO and HOMO eigenvalues and as the differences of the band-gap
edges at the level 0.5 eV$^{-1}$ in EDOS normalized so that the maximum value 
within the valence band is equal to unity.}
\label{tab:bandgaps}
\end{table}

Localization of the electronic states near the optical band-gap edges
is of great interest for studies of photoinduced phenomena.
A close-up of the inverse participation ratios at energies near the
band edges for models 1-4 is shown in Fig.~\ref{fig:eprs}.
If topological (or coordination) defects are present in a model, as in
model 1, some of the states, especially at the bottom of the
conduction band, are localized at them (see Fig.~\ref{fig:eprs}(a)).
This tendency was also emphasized in Ref.~\onlinecite{Drabold_PRB_00}.
A comparable, or even greater degree of localization, is observed in
model 2 (see Fig.~\ref{fig:eprs}(b)) where there are no coordination
defects, but there is an appreciable concentration of homopolar bonds
(or chemical defects).
As the number of chemical defects goes to zero from model 2 to
model~4, it is seen that localization at the top of the valence band
remains qualitatively similar, whereas at the bottom of the conduction
band it becomes less pronounced (see Fig.~\ref{fig:eprs}(b)-(d)).

The general picture is that, at the top of the valence band, the
eigenstates are predominantly localized at what can be called
sulphur-rich regions, where several sulphur atoms are closer than
about 3.45~\AA, i.e.  their interatomic distances are on the low-$r$
side of the second peak in $g_{\mathrm{S-S}}(r)$ shown in
Fig.~\ref{fig:pdf200d}(c) or some of these atoms form homopolar S-S
bonds.
For instance, most of the HOMO level in model 1 is localized at two
sulphur atoms separated by 3.42~\AA~ and which are part of the
molecule-like fragment depicted in Fig.~\ref{fig:two5rings}.
By inspecting the projected (local) IPRs in
Fig.~\ref{fig:ledos200d}(b) at the optical gap edges, it is seen that
the IPRs are greatest for the S atoms.
It appears that the localization at the top of the valence band is
facilitated by the proximity of the lone-pair p orbitals in the
sulphur-rich regions.

At the bottom of the conduction band, the states tend to localize at
four-membered rings in all models, and at S-S homopolar bonds (some of
these bonds are in five-membered rings) and IVAPs when such defects
are present.
In model 4, all three conduction-band states with an IPR greater than
0.025 (see Fig.~\ref{fig:eprs}(d)) are localized at four-membered
rings.

\section{Conclusion}

We have generated several models of amorphous arsenic sulphide by
using a density-functional-based tight-binding method.
All models agree very well with the neutron-scattering experimental 
structural data.
We observe a tendency for formation of quasi-molecular structural
groups which suggests that amorphous chalcogenides can be viewed as
nanostructured materials.
Vibrational properties are also in agreement with experimental
results.

In models containing both homopolar bonds and topological defects, a
significant degree of electronic-state localization has been observed
near both band-gap edges.
Although the coordination-number defects are optically active, their
presence may not be necessary for exhibiting photostructural changes
when there is a sufficient concentration of homopolar bonds in the
material.
This statement is supported by the observation that, upon removal of
the coordination defects from the system, the degree of
electronic-state localization is not reduced in the resultant
continuous network model with homopolar bonds.
Furthermore, the valence-alternation defect concentration is estimated
to be rather small (10$^{17}$ cm$^{-3}$ in
Ref.~\onlinecite{Feltz_AIMG}) compared to the atomic density of
about $2\times10^{25}$~cm$^{-3}$.
This indicates that, in the volume occupied by our 200-atom models, it
is less likely to find a topological defect than to come across none.

A stoichiometric continuous network model has allowed us to identify
the structural motifs where electronic eigenstates predominantly
localize at the optical band edges, in the absence of coordination and
chemical defects.
These are sulphur-rich regions for the top of the valence band and
four-membered rings for the bottom of the conduction band.
Electronic properties of this glass model with all-heteropolar bonds
are very similar to those of the corresponding crystalline phase,
orpiment, most notably the clean gaps in both s and p bands.
Although the valence band in this case contains about as many
localized states as in models with defects, the conduction band has
very few of them.
It is expected that, if there were no four-membered rings in this
structure, all states in the conduction band would be virtually
delocalized.

Therefore we conclude that perhaps the dominant contribution to
photo-induced effects originates from the presence of electronic
states localized in the vicinity of homopolar bonds, in support of the
theoretical models where the photo-induced structural changes are
attributed to the presence of homopolar bonds in these materials (see,
e.g., Refs.~\onlinecite{Kolobov_PIMAS,Spotnyuk_2003}).
Although electronic states can also localize in all-heteropolar
regions and in the vicinity of the topological defects, the
contribution of such states is likely to be rather small due to the
low degree of localization in the conduction band and the low
concentration of such defects, respectively.
Verification of this conjecture requires excited-state calculations 
and is beyond the scope of the present paper.
We plan to do these calculations in the future.

\section*{Acknowledgments}

S.I.S. is grateful to the EPSRC for financial support.  We thank the
British Council and DAAD for provision of financial support.


\end{document}